\documentclass[sigconf]{acmart}

\usepackage{pdflscape,array,booktabs}

\begin{document}

\title{Towards Code Summarization of APIs Based on Unofficial Documentation Using NLP Techniques}

\author{AmirHossein Naghshzan}
\email{amirhossein.naghshzan.1@ens.etsmtl.ca}
\affiliation{%
  \institution{École de Technologie Supérieure}
  \streetaddress{1100 Notre-Dame St W}
  \city{Montreal}
  \state{Quebec}
  \country{Canada}
  \postcode{H3C 1K3}
}

\settopmatter{printacmref=false}

\begin{abstract}
Each programming language comes with official documentation to guide developers with APIs, methods, and classes. However, in some cases, official documentation is not an efficient way to get the needed information. As a result, developers may consult other sources (e.g., Stack Overflow, GitHub) to learn more about an API, its implementation, usage, and other information that official documentation may not provide. 

In this research, we propose an automatic approach to generate summaries for APIs and methods by leveraging unofficial documentation using NLP techniques. Our findings demonstrate that the generated summaries are competitive, and can be used as a complementary source for guiding developers in software development and maintenance tasks.
\end{abstract}

\keywords{natural language processing, code summarization, unsupervised learning, deep learning, transformers}

\maketitle

\section{Research Problem and motivation}
During software development steps, programmers need sources to gain information about different APIs, methods, or classes to become familiar with them and understand how to use/implement them. Despite the fact that developers mostly rely on official documentation as their primary source of information, previous studies have shown that official documentation is not always the best way of extracting information since it may be long and time-consuming and even in some cases lacks completeness~\cite{ponzanelli2015turning}. Therefore, most developers use unofficial documents like StackOverflow, GitHub, and other sources to get the needed information for their tasks. 

To address this problem, we decided to automatically generate summaries for APIs discussed on informal documents based on the natural languages around them and their surrounding context. For this research, we selected StackOverflow as unofficial documentation and Android programming language for studying its APIs and methods.

\section{Related Work}
Automatic source code summarization has received attention recently since it is an interesting topic for researchers. However, they were mostly concerned with producing a summary for a code block, not an API, and also very small number of them used unofficial documentation. Here we focus on the most relevant research to our study.

 Sridhara {\em et al.}~\cite{Sridhara} have used Natural Language Processing to extract information to summarize Java methods. The result of their work demonstrated that the comments generated for Java methods from source code are accurate and do not miss significant information in the process of creating comments. 
 
 Moreno {\em et al.}~\cite{Moreno} have introduced a tool, JStereoCode, that can automatically define stereotypes for methods and classes using Java source code. The main advantage of this approach is that developers can use it as a recovery for their projects.
 
 To understand the use of the Java method, Hu et al.~\cite{Hu} proposed a new approach to generate code comments from a large code corpus called DeepCom. In their approach, by applying Natural Language Processing (NLP) techniques some limitations such as extracting accurate keywords from the methods that are not named correctly are solved. 

Mohammed {\em et al.}~\cite{Mohammed} introduced a technique that covers both relating high-level explanations to the particular areas of the source code that implement them and communicating how parts of the code interact through control and data dependencies, automatically. 

\section{Methodology}

\begin{table*}
  \centering
\caption{Examples of generated summaries for onCreate().}
\renewcommand{\arraystretch}{1.7}
\begin{tabular}{m{2.5cm} m{11cm}}
 \hline
 \centering{\textbf{Algorithm}}& \multicolumn{1}{c}{\textbf{Summary}} \\
 \hline
 \centering{TextRank}   & 
When the app is launched,  the first thing that's going to run is onCreate()  in this case,  onCreate()  has a  method that inflates the view of your activity, that method is called setContentView(). When you pass data from one activity to another using a  Bundle,  the data is received inside the onCreate()  method of the second activity, not insideonActivityResult() unless you've specifically implemented that.\\
    \hline
 \centering{GPT-3}  & 
OnCreate is the first method in the Activity lifecycle whose job is to create an activity. It is called the first time the Activity is created, or when there is a configuration change like a screen rotation. This is the beginning state of an activity lifecycle in which its view should be set. Most of the activity initialization code goes here.\\
    \hline
\end{tabular}
  \label{table:result}
\end{table*}

Our research methodology consists of three main steps: data collection, building a corpus, and summarization.

\subsection{Data Collection}
The first step of our research is data collection. Using Stack Exchange API, we extracted all StackOverflow's questions having {\em Android} tag from January 2009 till
April 2020. After that, we extracted all answers to those questions and ended up with 3,084,143 unique Android posts which formed our dataset. To extract mentioned APIs in posts, we used {\em Code Snippet} feature of StackOverflow. This feature highlights code blocks in posts to make it easier for users to find them by adding an HTML code tag (<code>) around them. After extracting all code blocks, using regex patterns we filtered the code blocks that had specific characterization like having multiline, spaces, etc. to find those code blocks that are API and method names.

\subsection{Building a Corpus}
The second step is to build a corpus. Not all the answers in StackOverflow contain accurate information. So we removed the answers with a score below the average score of all Android answers which is 2. In the remaining answers, For each API we tried to collect the relevant sentences of that answer to add them to our corpus. For this purpose, we extracted three sentences for each detected API. One sentence that the API appeared in, the previous sentence, and the next sentence. We believe these sentences have the highest chance of being relevant to the APIs. Moreover, we added the first sentence of those answers to our corpus since our observations proved the first sentence usually contains important information.

\subsection{Summarization}
After building the corpus and doing cleaning and pre-processing works using the standard Natural Language Toolkit (NLTK), the next step is the summarization task. For this step, we tried two different approaches. In our recent research~\cite{Naghshzan} we used the TextRank algorithm which is an unsupervised learning method for generating extractive summaries based on the given corpus~\cite{textrank}. We selected this simple algorithm to focus on validating the feasibility of our approach. As mentioned in the paper~\cite{Naghshzan} the results were promising so we decided to continue our work by improving the quality of summaries. For this purpose, as the next step, we used deep learning algorithms to generate abstractive summaries. Therefore we replace the TextRank algorithm with GPT-3 which is a transformer technique in NLP~\cite{gpt3}. Since TextRank is an extractive algorithm, the generated summaries contain the exact sentence that appeared in StackOverflow and some cases even not relevant to the next sentence. However, changing the algorithm to an abstractive one, increased the quality of summaries and made it more coherent. The sentences Table \ref{table:result} shows the results of summarization for app.activity.onCreate which is the most repeated Android method in StackOverflow.

\section{Empirical Evaluation}

To learn more about how developers see the accuracy of our generated summaries, we conducted an empirical study. We have used a completely randomized design~\cite{design} with 16 professional Android developers. We provided TextRank's generated summaries and asked them to read and evaluate the accuracy, coherence, length, and usefulness of summaries using the Likert scale. Moreover, we provided the official Google documentation and asked them to compare our generated summaries with official descriptions in terms of usefulness.

The results have shown that our summaries are valuable sources for software engineers during their development tasks. The main findings of our empirical evaluation are as follows:

\begin{itemize}

\item The length of summaries is adequate, according to all developers included in this study (100\%).

\item In terms of coherence, about half of the participants (58\%) agree that the generated summaries are coherent.

\item The majority of participants (73\%) agreed that our summaries include reliable information concerning Android APIs.

\item Participants agreed that an integrated plugin to display our automatically generated summaries would be useful, with a rating of 4.1 out of 5.

\end{itemize}

Furthermore, as an important finding, while participants mostly preferred the official descriptions, they believe that the created summaries may be utilized as a complementary for official documents as they contain information that is not discussed in the official documentation~\cite{Naghshzan}.

\section{Conclusion and Future Work}

In this research, we proposed a novel approach to generate summaries for APIs and methods discussed on unofficial documentation based on their surrounding context. We selected StackOverflow Android posts as our dataset and applied TextRank and GPT-3 algorithms as the main methods of summarization. Finally, we conducted an empirical evaluation with software developers to evaluate our generated summaries. 

Our empirical evaluation shows that our generated summaries are useful for development tasks and more importantly, participants agreed that these summaries could be used as a complementary for official descriptions.

In our future work, we are interested in comparing our approach with state-of-the-art methods. In addition, we will conduct a large-scale empirical evaluation to check if our summaries are useful during real development tasks. Finally, we are interested in integrating our approach as a plugin for IDE applications to help software engineers during development processes.



\end{document}